\documentclass[dvieps]{acta}
\usepackage{supertabular,lscape,epsfig}
\usepackage{amssymb}
\usepackage{amsmath}
\usepackage{siunitx}

\SetPages{0}{0}

\SetVol{00}{0000}

\begin{document}

\begin{Titlepage}
\Title{A comprehensive study of the eclipsing binaries {V1241~Tau} and {GQ~Dra}}
% \Author{A~u~t~h~o~r~1, First Initials.,}{Affiliation1\\
% e-mail:}

% \Author{A~u~t~h~o~r~N, First Initials.}{AffiliationN\\
% e-mail:}

% or 

% \Author{A~u~t~h~o~r1, First Initials. \dots and A~u~t~h~o~r~N, 
% First Initials.}{Common Affiliation\\
% e-mail:}

% or 

\Author{B.~U~l~a~\c{s}$^{1,2}$, K.~G~a~z~e~a~s$^3$, A.~L~i~a~k~o~s$^4$, C.~U~l~u~s~o~y$^5$, I.~ S~t~a~t~e~v~a$^6$, N.~E~r~k~a~n$^{7,2}$,  M.~N~a~p~e~t~o~v~a$^6$ and I.~Kh.~I~l~i~e~v$^6$}
%{$^1$\.{I}zmir Turk College Planetarium, 8019/21 sok., No: 22, \.{I}zmir, Turkey
{$^1$Department of Space Sciences and Technologies, Faculty of Arts and Sciences, \c{C}anakkale Onsekiz Mart University, Terzio\v{g}lu Campus, TR-17100, \c{C}anakkale, Turkey\\
e-mail:burak.ulas@comu.edu.tr\\
$^2$\c{C}anakkale Onsekiz Mart University, Astrophysics Research Centre and Ulup{\i}nar Observatory, TR-17100, \c{C}anakkale, Turkey\\
$^3$Section of Astrophysics, Astronomy and Mechanics, Department of Physics, National and Kapodistrian University of Athens, GR-157 84, Zografos, Athens, Greece\\
%e-mail:\\
$^4$Institute for Astronomy, Astrophysics, Space Applications and Remote Sensing, National Observatory of Athens, GR-15236, Penteli, Athens, Greece\\
$^5$Department of Physics and Astronomy, Botswana International University of Science and Technology (BIUST), Plot 10017, Private Bag 16, Palapye, Botswana\\
$^6$Institute of Astronomy and NAO, Bulgarian Academy of Sciences, Sofia 1784, Bulgaria\\
$^7$Department of Physics, Faculty of Arts and Sciences, \c{C}anakkale Onsekiz Mart University, Terzio\v{g}lu Campus, TR-17100, \c{C}anakkale, Turkey}
%e-mail:}

\Received{Month Day, Year}
\end{Titlepage}

\Abstract{We present new photometric and spectroscopic observations and analyses for the eclipsing binary systems {V1241~Tau} and {GQ~Dra}.  Our photometric light and radial velocity curves analyses combining with the TESS light curves show that both are conventional semi--detached binary systems. Their absolute parameters are also derived. We present the $O-C$ analyses of the systems and we propose the most possible orbital period modulating mechanisms. Furthermore, Fourier analyses are applied to the photometric residual data of the systems to check for the pulsational behavior of the components.  We conclude that the primary component of the system {GQ~Dra} is a $\delta$~Sct type pulsator with a dominant pulsation frequency of 18.58~d$^{-1}$ based on our $B$ filter residual light curve although it can not be justified by 30-minute cadence TESS data. No satisfactory evidence of pulsational behaviour for {V1241~Tau} was verified. Finally, the evolutionary tracks of the components of both systems are calculated, while their locations within evolutionary diagrams are compared with other Algol-type systems.}{binaries: eclipsing -- stars: oscillations (including pulsations) -- stars: fundamental parameters -- stars: evolution -- stars: individual:V1241~Tau, GQ~Dra}

\section{Introduction}
Our knowledge on the properties of the pulsating components in binary star systems has been increased dramatically within the last twenty years. 69 out of 636 $\delta$~Sct stars in the catalog of Rodr{\'\i}guez~{\it et al.} (2000) were also appeared in the catalog of the components of double and multiple stars of Dommanget and Nys (1994). Mkrtichian~{\it et al.} (2002) recommended a new approach for determining the evolutionary properties of pulsating stars in Algol type binary systems by inspecting six {\it oEA} (Oscillating Eclipsing Systems of Algol type) stars. The effect of mass transfer on the pulsational characteristics of the oscillating components was also probed by Mkrtichian~{\it et al.} (2003). Soydugan~{\it et al.} (2006) remarked about the possibility of a relation between pulsational and orbital periods for the systems having $\delta$~Sct type primary components. Pulsational and binary properties of 74 systems with $\delta$~Sct stars was presented in detail by Liakos~{\it et al.} (2012). Changs~{\it et al.} (2013) published the statistical features of $\delta$~Sct stars within the Milky Way galaxy and noted that 141 of them were reported as companions of binary systems in previous double star catalogs. The catalog of Liakos and Niarchos (2015) lists the pulsational and binarity properties of more than a hundred systems with $\delta$~Sct components. Liakos and Niarchos (2017) catalogued 199 binary systems having $\delta$~Sct components and discussed their demographics and orbital-pulsational periods relation.

The studies on eclipsing binaries with mass accreting pulsating components open new doors in understanding the inner, dynamical, and evolutionary properties of binary systems. Mkrtichian~{\it et al.} (2002) indicated that the splittings of low- and high-degree non-radial pulsations are the keys in the estimation of internal and differential rotations of the stars. They also highlighted that the change in the pulsation period is a useful tool to predict the evolutionary stage of the mass-gainer component. Mkrtichian~{\it et al.} (2003) remarked that the mode splitting can be used to determine the accurate rotation period and therefore helps to hold a view about the asynchronicity of binary systems. Moreover, they discussed the relation between density variation due to mass accretion and the pulsational period change. This connection allows us to estimate mass accretion which plays a vital role in binary star evolution.  

There are plenty of remarkable cases among binary systems with pulsating components which were investigated by researchers so far. Skarka~{\it et al.} (2019) denoted that HD~99458 is an eclipsing binary system which consists of an Ap star primary showing $\delta~Sct$ type pulsating characteristics and an M-dwarf secondary. A $\delta~Sct$ star in EL~CVn type eclipsing binary system was also investigated by Wang~{\it et al.} (2018) who indicated that the secondary is a pre-He white dwarf. A similar case was also delved by Guo~{\it et al.} (2017) for the EL~CVn type system KIC~8262223. The authors underlined the p-mode pulsations of the primary component which is excessively affected by the mass transfer from the secondary companion. Another notable system is KIC~3958884. Maceroni~{\it et al.} (2014)  remarked that the primary component of the binary is a hybrid pulsator having high order g-modes excited by phenomena related to binarity effects.

Pickering (1908) remarked that the light variation of {V1241~Tau} was discovered by Henrietta S. Levaitt. Gaposchkin (1953) noted that the system is an eclipsing binary. $ubvy$ and H$_{\beta}$ photometry of the system was obtained by Drilling and Pesch (1973). Sarma and Abhyankar (1979) observed it during 21 nights in $UBV$ filters. They applied a period analysis to their times of minima and amended the orbital period value. The authors analysed the light curve and concluded that {V1241~Tau} is a detached binary system. The light curves of Sarma and Abhyankar (1979) were also analysed by Giuricin and Mardirossian (1981). They found that the configuration of the system is detached, but close to be in contact, and claimed that the spectral type of the secondary component is about K0. Russo and Milano (1983) concluded that the system is a typical Algol, from an evolutionary point of view, by analysing the observations of Sarma and Abhyankar (1979). Srivastava and Kandpal (1986) presented the $BV$ light curves of the system and calculated an improved ephemeris. Using $B-V$ indices, they also derived that the primary and secondary components are A5 and K0 stars, respectively. Arentoft~{\it et al.} (2004) analysed their own light curve using two different mass ratio values and they obtained three times of minimum. The system was in the target list of ASAS (All Sky Automated Survey, Pojmanski 2003). Kazarovets~{\it et al.} (2006) listed suggested renamings for early miscoordinated variable stars. {WX~Eri} was renamed as {V1241~Tau} in their table. Yang~{\it et al.} (2012) presented a new light curve solution for their and ASAS data and concluded that the system is a near--contact binary (NCB). They also mentioned a period variation which could be caused by a third body. The system also appeared in $uvby\beta$ catalog of Paunzen (2015).

The possible pulsating behaviour of the primary component of {V1241~Tau} was firstly mentioned by Sarma and Abhyankar (1979). They pointed out that the primary component could be a $\delta$~Sct-type pulsator by virtue of wave-like deviations of the observed points from the theoretical light curve. Rodr{\'\i}guez~{\it et al.} (2000) included the star in their $\delta$~Sct catalog.  However, Srivastava and Kandpal (1986) could not find any oscillations by analysing their light curve. Similarly,  Arentoft~{\it et al.} (2004) concluded that there is no evidence for a $\delta$~Sct-type variable component according to their analysis. Moreover, Pigulski and Michalska (2007) did not find any significant signal in the residuals of ASAS-3 light curve of the system. Zhang~{\it et al.} (2013) included the system in their study on the relation between pulsational and orbital periods of $\delta$~Sct stars in eclipsing binary systems. Liakos and Niarchos (2017) also listed it in their catalog of $\delta$~Sct stars in binaries by remarking that the presence of pulsations in the primary component is ambiguous. Recently, the system was referred by Ziaali~{\it et al.} (2019), who focused on the period-luminosity relation of $\delta$~Sct stars.

The variable behaviour of the system {GQ~Dra} was firstly discovered by Hipparcos (ESA 1997). Atay~{\it et al.} (2000) published photoelectric $BVR$ observations and obtained minima times. Qian~{\it et al.} (2015) modeled the light curve obtained by the Lunar-based Ultraviolet Telescope. They also analyzed the orbital period of the system and concluded that it shows increment, which could be a consequence of mass transfer from the secondary component to the primary. Liakos and Niarchos (2017) included the system in their catalog of $\delta$~Sct stars in binaries.

The efforts made by previous researchers indicated that these systems worth to be investigated deeply, especially in terms of binarity and pulsational characteristics. In the following section, we present details concerning our observations on these systems. The light and radial velocities curves models of both systems are given in Sec.~3. The orbital period studies are stated in Sec.~4. The investigation for the pulsational behaviour of the targets is presented in Sec.~5. The evolutionary tracks and the positions of the components of the systems as well as their comparison with other similar systems are presented in Sec.~6. Finally, the last section includes a summary of our results and our conclusions.

\section{New Observations}
\subsection{Photometry}

CCD photometric observations of {V1241~Tau} and {GQ~Dra} were carried out with the 0.40 m f/8 Cassegrain reflector and the SBIG ST-10XME CCD camera equipped with the $BVRI$ (Bessell specification) filters at the University of Athens Observatory (UOAO). Typical seeing was 2 arcsec during the observing periods.

V1241~Tau was observed in November 2012 within 6 nights. This resulted in 2647 data points in $BVRI$ filters; 700, 649, 649, 649 data points in each filter, respectively.
The exposure times were set as 60, 20, 15, and 20 seconds for $B$, $V$, $R$, and $I$ filters, respectively. The comparison and check stars used are GSC~4709:1022 ($V$=10$^m$.11, $B-V$=1$^m$.19) and GSC~4709:1211 ($V$=10$^m$.19, $B-V$=0$^m$.50). GQ Dra was observed between 22 May and 10 June 2016 within 16 nights. We obtained 3382, 3633, 3188, 3715 data points in $B$, $V$, $R$, and $I$ filters, respectively, thus 13912 points in total. The exposure times were 40 sec for $B$, 20 sec for $V$,$R$, and $I$ filters. The comparison and check stars were GSC~3521:0391 ($V$=9$^m$.90, $B-V$=1$^m$.14) and GSC 3520:1598 (10$^m$.90), respectively. We performed $BVRI$ series photometry for both target on each night of observations. The light curve data of both systems are given in Table~1.

\MakeTable{lccc}{12.5cm}{$BVRI$ light curve data of the systems. $V-C$ refer to differential magnitude, difference between magnitudes of variable and comparison stars. This table is for guidance regarding the electronic edition which is available online.}
{\hline
Name & Filter & HJD & $V-C$ \\
\hline
{V1241~Tau}& B& 2456234.35169& -1.447 \\
{V1241~Tau}& B& 2456234.35340& -1.443 \\
{V1241~Tau}& B& 2456234.35510& -1.459 \\
{V1241~Tau}& B& 2456234.35681& -1.438 \\
{V1241~Tau}& B& 2456234.35852& -1.447 \\
{~~~~~\ldots} & \ldots & \ldots & \ldots\\
\hline
}

\begin{sloppypar}
In the case of {V1241~Tau}, aperture photometry using {\tt IRAF~(DIGIPHOT.APPHOT)} package (Davis 1994) was applied. The photometric data of {GQ~Dra} were also reduced with aperture photometry technique, using the {\tt C-Pack} photometric package, which utilizes photometric routines of {\tt IRAF} (Hroch 1998). 
\end{sloppypar}

Our targets fall into the observing region of TESS (Transiting Exoplanet Survey Satellite; Ricker~{\it et al.} 2009, 2015) mission. Therefore, we include the systems' TESS light curves to our analyses (Sec.~3 and 5). The 2-minute cadence photometric time series covering 14287 data points are available for V1241~Tau through MAST portal\footnote{https://mast.stsci.edu/portal/Mashup/Clients/Mast/Portal.html}. The light curve of GQ~Dra was also extracted from the full-frame images (FFI) of the system taken with 30-minute cadence by TESS. The {\tt eleanor} (Feinstein~{\it et al.} 2019) tool was used during the extraction of images into 1115 data points of the light curve.

The multi--colour and TESS light curves of the systems are shown in Figs.~1 and 2. Calculated times of minimum from our $BVRI$ observations and TESS data are listed in Table~2.

\MakeTable{lccc}{12.5cm}{Times of minimum as derived from our observations and TESS light curves. This table is for guidance regarding the electronic edition which is available online.}
{\hline
Name & HJD & Error & Type \\
\hline
{V1241~Tau} & 2456235.43700 & 0.00030 & Sec \\
{V1241~Tau} & 2456242.43400 & 0.00010 & Pri \\
{V1241~Tau} & 2458411.33939 & 0.00009 & Sec \\
{V1241~Tau} & 2458411.75014 & 0.00001 & Pri \\
{V1241~Tau} & 2458412.16263 & 0.00007 & Sec \\
{~~~~~\ldots} & \ldots & \ldots & \ldots\\
\hline
}

\subsection{Spectroscopy}

Spectroscopic observations have been carried out with the Echelle Spectrograph Rozhen (ESpeRo) -- new-commissioned fiber-fed echelle spectrograph -- installed at the 2-m Ritchey-Chretien-Coude telescope of Rozhen National Astronomical Observatory (Bonev~{\it et al.} 2017). Andor iKon-L CCD-camera was used to record stellar spectra with a typical resolving power of about 35\,000 in the wavelength range between 3850~\AA\,and 9000~\AA. 

We have taken into account the time resolution requirements in order to have a good S/N ratio which means longer exposures but also not too much to reduce the actual radial velocity variations. Following these limitations we chose the exposure time to be 30~min. This resulted in an S/N of about 40 in phase coverage less than 0.025 for V1241 Tau and 0.03 for GQ Dra respectively. Before and after the exposures on the  target systems, standard stars spectra were also obtained and used for the subsequent wavelength calibration.%The observing procedure was following thar-object-thar sequence. 

IRAF standard procedures were used for bias subtracting, flat-fielding and wavelength calibration. The wavelength calibration was made for every thar-spectrum and at the end the mean wavelength solution from both thar-spectra was taken. The final spectra were corrected to the heliocentric wavelengths.

%After the calculation of the atmospheric parameters $T_{eff}$ and $logg$,The spectra were taken during the secondary minimu 
For the calculation of the atmospheric parameters $T_{eff}$ and $logg$, we chose the spectra taken during the secondary minimum phase ($\sim$0.5) which the primary dominates the spectrum almost totally and the contribution from the secondary is negligible. Model atmospheres were calculated under ATLAS\,12 code. The VALD atomic line database (Kupka~{\it et al.} 1999), which also contains Kurucz (1993) data, was employed to create a line list for the synthetic spectra. The code {\tt SYNSPEC} (Hubeny~{\it et al.} 1994, Krti{\v{c}}ka 1998) was used to calculate the synthetic spectra. A microturbulence of 2\,km\,s$^{-1}$ was assumed.

The best fit of H$\beta$ and H$\alpha$ lines for V1241~Tau was obtained for the atmosphere model with $T_{\rm eff}=7100\pm 100~{\rm K}$ and $\log g=4.0$. The spectral line of Mg\,II $\lambda$\,4481\,\AA\,was used for the determination of the projected rotational velocity. The match between the synthetic and observed profile gave as a result $v\sin\,i=100~{\rm km\,s^{-1}}$.
The results for GQ~Dra were: $T_{\rm eff}=8750\pm 100~{\rm K}$, $\log g=4.0$, $v\sin\,i=90~{\rm km\,s^{-1}}$. 

The obtained projected rotational velocities $v\sin\,i$ of both systems were too high and prevented the correct measurements of single spectral lines. That was the reason for measuring the radial velocities by using the procedure FXCOR of IRAF which uses Fourier cross-correlation between the observed and template spectra. Synthetic spectra of selected spectral regions were calculated to be used as templates for this procedure.

Since the stars in our systems rotate fast - i.e., the projected rotational velocities are close to 100~km~s$^{-1}$ and all the lines are weak and broad, we could not detect any components, even in Balmer lines, during the orbital period cycle. The lines appear as single without any splitting. Therefore, we conclude that the secondary components of the system, are very faint and they do not show any impact on the line profiles. The observed radial velocities for the primary component of each system ($RV_1$) are given in Table~3 and they are plotted in the bottom panels of Figs.~1 and 2.

\MakeTable{lr}{12.5cm}{Observed radial velocities for both systems.}
{\hline
HJD-2400000 & $RV_1$~(km~s$^{-1}$)               \\
\hline
\multicolumn{2}{c}{V1241~Tau}\\
\hline
56226.47363&62\,$\pm$\,14\\
56227.46216&113\,$\pm$\,13\\
56231.35810&40\,$\pm$\,15\\
56732.24383&93\,$\pm$\,34\\
56735.24020&67\,$\pm$\,15\\
56736.23320&105\,$\pm$\,11\\
56737.23483&-17\,$\pm$\,25\\
56738.23307&-90\,$\pm$\,17\\
57376.22666&-85\,$\pm$\,3\\
57376.24806&-88\,$\pm$\,1\\
57376.27172&-92\,$\pm$\,1\\
57377.22202&-50\,$\pm$\,2\\
57377.24370&-38\,$\pm$\,3\\
57377.26504&-23\,$\pm$\,2\\
57377.43934&96\,$\pm$\,2\\
57377.46076&103\,$\pm$\,3\\
57380.28879&-58\,$\pm$\,3\\
\hline
\multicolumn{2}{c}{GQ~Dra}\\
\hline
57493.46159&43\,$\pm$\,7\\
57493.48531&62\,$\pm$\,5\\
57494.44778&62\,$\pm$\,6\\
57494.47123&52\,$\pm$\,19\\
57494.49469&40\,$\pm$\,1\\
57495.46601&-70\,$\pm$\,9\\
57495.48943&-76\,$\pm$\,7\\
57495.51288&-77\,$\pm$\,3\\
57522.48494&1\,$\pm$\,11\\
57522.50884&16\,$\pm$\,6\\
57523.48002&88\,$\pm$\,12\\
57523.50387&90\,$\pm$\,27\\
57524.47677&-29\,$\pm$\,9\\
57555.55017&66\,$\pm$\,9\\
57556.43253&75\,$\pm$\,11\\
\hline}

\section{Photometric and Spectroscopic Analyses}
The light curves of both systems were analysed using the {\tt PHOEBE} (Pr{\v{s}}a and Zwitter 2005) software, which utilizes the Wilson-Devinney code (Wilson and Devinney 1971). The inclination $i$, the temperature of the secondary component $T_2$, the mass ratio $q$, the velocity of the centre of mass $V_0$, the surface potential value $\Omega_1$, luminosity of the primary component $L_1$ were set as adjustable parameters during the analyses. The third light parameter $l_3$ was left free during the analysis following the previous researchers and the findings from our $O-C$ analyses (Sec.~4). However, we could not find any meaningful resulting values, hence it was neglected. Gray and Nagel (1989) remarked that the granulation boundary for main sequence stars is placed close to F0 spectral type, which corresponds to a temperature value of 7460 K (Cox 2000). Therefore, we adopted the gravity darkening coefficients $g$ and albedos $A$ as given in Lucy (1967) and Rucinski (1969). The limb darkening coefficients $x$ were taken from the tables of van Hamme (1993) for $BVRI$ light curves and from Claret (2017) for the TESS data.

The initial mass ratio values for the analyses were derived as follows: Initially, we estimated the masses of the primaries in $\pm0.3~$M$_{\odot}$ interval (M$_{1min}$=1.2~M$_{\odot}$, M$_{1max}$=1.8~M$_{\odot}$ for {V1241~Tau} and M$_{1min}$=2.1~M$_{\odot}$, M$_{1max}$=2.7~M$_{\odot}$ for {GQ~Dra}) by taking into consideration the temperatures (T$_1$) and the amplitudes of the radial velocities curves (K$_1$). Then, we derived the minimum and maximum possible mass ratio values (0.47~$\leq~q~\leq$~0.56 for {V1241~Tau} and 0.31~$\leq~q~\leq$~0.34 for {GQ~Dra}) that correspond to the aforementioned mass intervals using the {\tt ABsParEB} software (Liakos 2015). The final mass ratio values were determined using the minimum squared residuals method $\Sigma(res)^{2}_{min}$ in the simultaneous analysis of light and radial velocities curves. More details for each system can be found in the following subsections.

%The simultaneous solution of the light and radial velocities curves of the systems was resulted with high values of $\Sigma(res)^{2}$ and unrealistic estimation for the errors. Therefore, 
In order to achieve a better fit, a more realistic estimation for the errors and, thus, a better removal of the binary model, which is crucial in further search for the pulsational properties, we analysed $B$, $V$, $R$ and $I$ and TESS light curves separately by combining each of them with the radial velocities data. The final values for the parameters were determined by calculating the average of the respective values of the individual models, while the errors are the standard deviations of them. %Furthermore, the standard deviations were adopted as the measure of dispersion in the resulting parameters as listed in Table~\ref{tablc}.

\subsection{{V1241~Tau}}
%The analysis was applied on more than 2500 photometric data points taken in $BVRI$ filters. The 2 minute cadence photometric time series are also available for the system thorough MAST portal\footnote{https://mast.stsci.edu/portal/Mashup/Clients/Mast/Portal.html}. Therefore, we include the TESS light curve, which covers 14287 data points, to our analyses. 
Because of the various suggestions in the literature (see Sec.~1) regarding the geometrical configuration of {V1241~Tau}, we tried to model the light curve assuming two different configurations: (i) semi--detached and (ii) detached. The model using the detached configuration was interrupted by the warning that the secondary component fills its Roche Lobe. Therefore, the analysis was continued for the conventional semi--detached configuration. Following our spectroscopic results, the effective temperature of the hotter star was set to 7100~K during the iterations. The radial velocities curve amplitude was calculated as $K_1$= 101$\pm$10~km~s$^{-1}$ by fitting the $RV_1$ data via sine function and by using $V_0$ and $i$ as derived from the analysis. The agreements between the theoretical light and radial velocities curves and the observations are drawn in Fig.~1, while the modeling parameters are listed in Table~4. Based on these results, the absolute parameters of the system were derived and they are given in Table~5.

\begin{figure}[htb]
\centering
\includegraphics[scale=0.85]{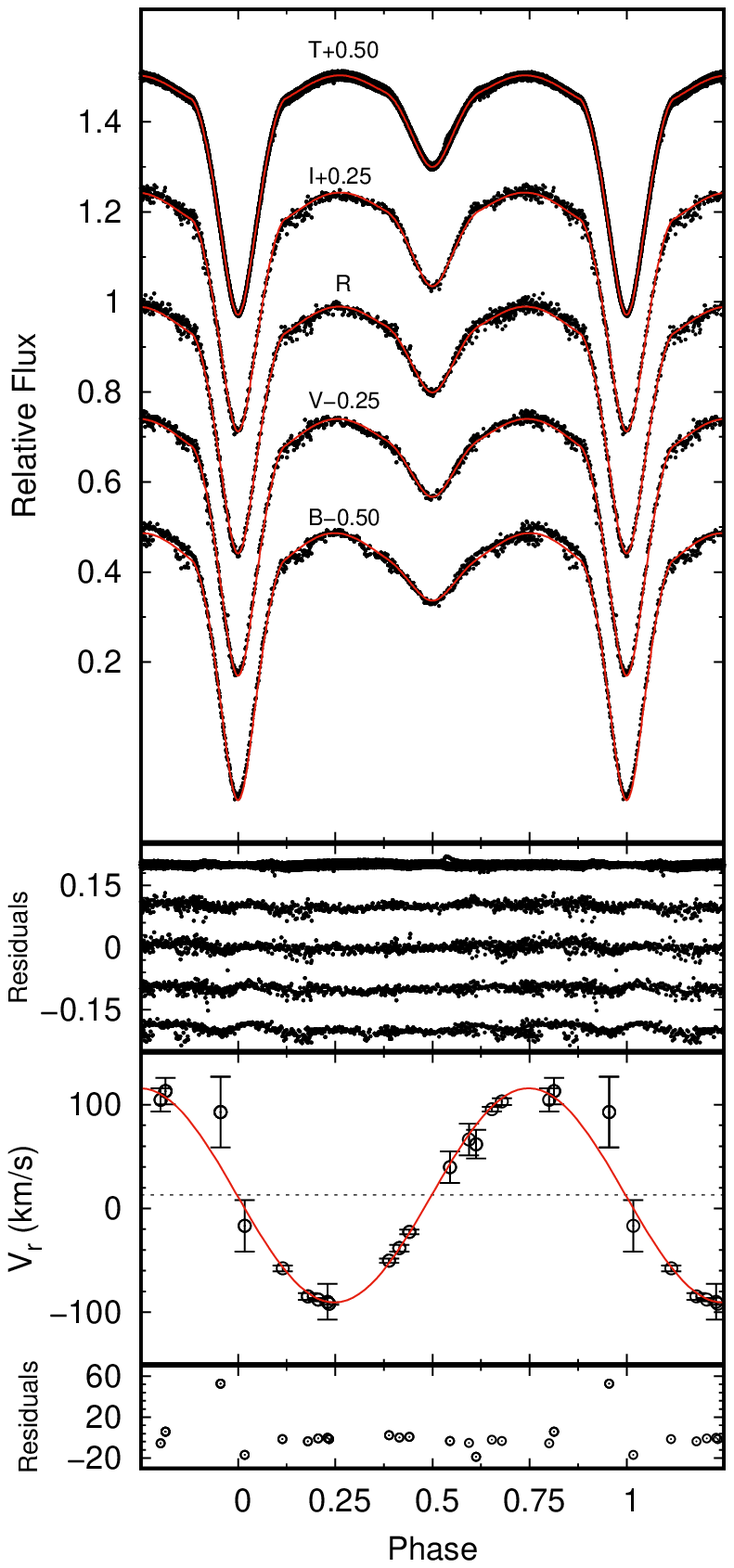}
\FigCap{Observed (points) and theoretical (lines) light curves of V1241~Tau and the residuals after subtraction of binary model from $T$, $I$, $R$, $V$, $B$ observations (two uppermost panel). Radial velocities curves and their residuals are plotted in the two lowermost panel. $T$ stands for TESS bandpass covering 600-1000~nm.}
\label{figlcs}
\end{figure}

\MakeTable{lcc}{12.5cm}{Results of light and radial velocities curves analyses for both systems. The standard deviations, 3$\sigma$, that corresponds to the last digit(s) are given in parentheses alongside values. T refers to the TESS bandpass.}
{\hline
Parameter                                   & V1241~Tau & GQ~Dra \\
\hline
$i$ (deg)                                   & 79.8(2)   &71.8(5)\\
$q$                                         & 0.481(5)   &0.320(5) \\
$V_0$~(km~s$^{-1}$)                         & 12.8(1)   &5.2(2) \\
%$K_1$~(km/s)                                & 99.6     & 81.2 \\
$T_1$~(K)                                   & 7100(100)      &8750(100) \\
$T_2$~(K)                                   & 4901(252)  &5109(281)\\
$\Omega _{1}$                               & 3.43(6)    &2.82(1)     \\
$\Omega _{2}=\Omega _{RL} $                 & 2.839 &  2.510      \\
%$\Phi$										&	--		&0.0031(1) \\
Fractional radius of primary                & 0.352(8)  &0.413(3) \\
Fractional radius of secondary              & 0.323(3)  &0.284(1) \\
Luminosity ratio $\frac{L_1}{L_1 +L_2}$:&   \\
$B$                                         & 0.93(2)  &0.968(6)      \\
$V$                                         & 0.88(2)  &0.961(5)      \\
$R$                                         & 0.85(2)  &0.945(6)       \\
$I$                                         & 0.82(2)  &0.918(5)       \\
$T$                                         & 0.83(4)  &0.931(2) \\
$x_1$, $x_2$:                                 &  \\       
$B$                                         & 0.786, 0.852  &0.767, 0.853      \\
$V$                                         & 0.687, 0.805  &0.661, 0.799       \\
$R$                                         & 0.589, 0.722  &0.535, 0.712       \\
$I$                                         & 0.496, 0.626  &0.422, 0.615       \\
$T$                                         & 0.547, 0.682  & 0.479, 0.659 \\
$g_1$, $g_2$                                & 0.32, 0.32  &1.0, 0.32      \\
A$_1$, A$_2$                                 & 0.5, 0.5    &1.0, 0.5      \\
Spot parameters:&  & \\
$\beta$ (deg)& -- &  50\\
$\lambda$ (deg)& -- &  275\\
$r$ (deg)&--  &  25 \\
$T_{f}$& -- &  0.85\\
$\chi^{2}$ & 3581.3 & 1062.5 \\
\hline}

\subsection{{GQ~Dra}}

%The number of data points for $BVRI$ light curves analyses of {GQ~Dra} was 13918. An additional light curve was also extracted from the full-frame images (FFI) of the system taken with 30 minute cadence by TESS satellite. The {\tt eleanor} (Feinstein~{\it et al.} 2019) tool was used during the extraction of images into 1115 data points.

We followed the same method similar to that of {V1241~Tau} along the analyses. The effective temperature of the primary component was assumed 8750~K following our spectroscopic results. $K_1$ was calculated as 72$\pm$10~km~s$^{-1}$. The slight difference between the two maxima of the $BVRI$ light curves led us to assume a cool spotted area on the surface of the secondary component during their analyses. The parameters of the spotted area are the co-latitude $\beta$, the longitude $\lambda$, the radius $r$ and the temperature factor $T_f$. The observed and the calculated light and radial velocities curves are drawn in Fig.~2. The results of the analysis are shown in Table~4, while the absolute parameters are listed in Table~5.

\begin{figure}[htb]
\centering
\includegraphics[scale=0.85]{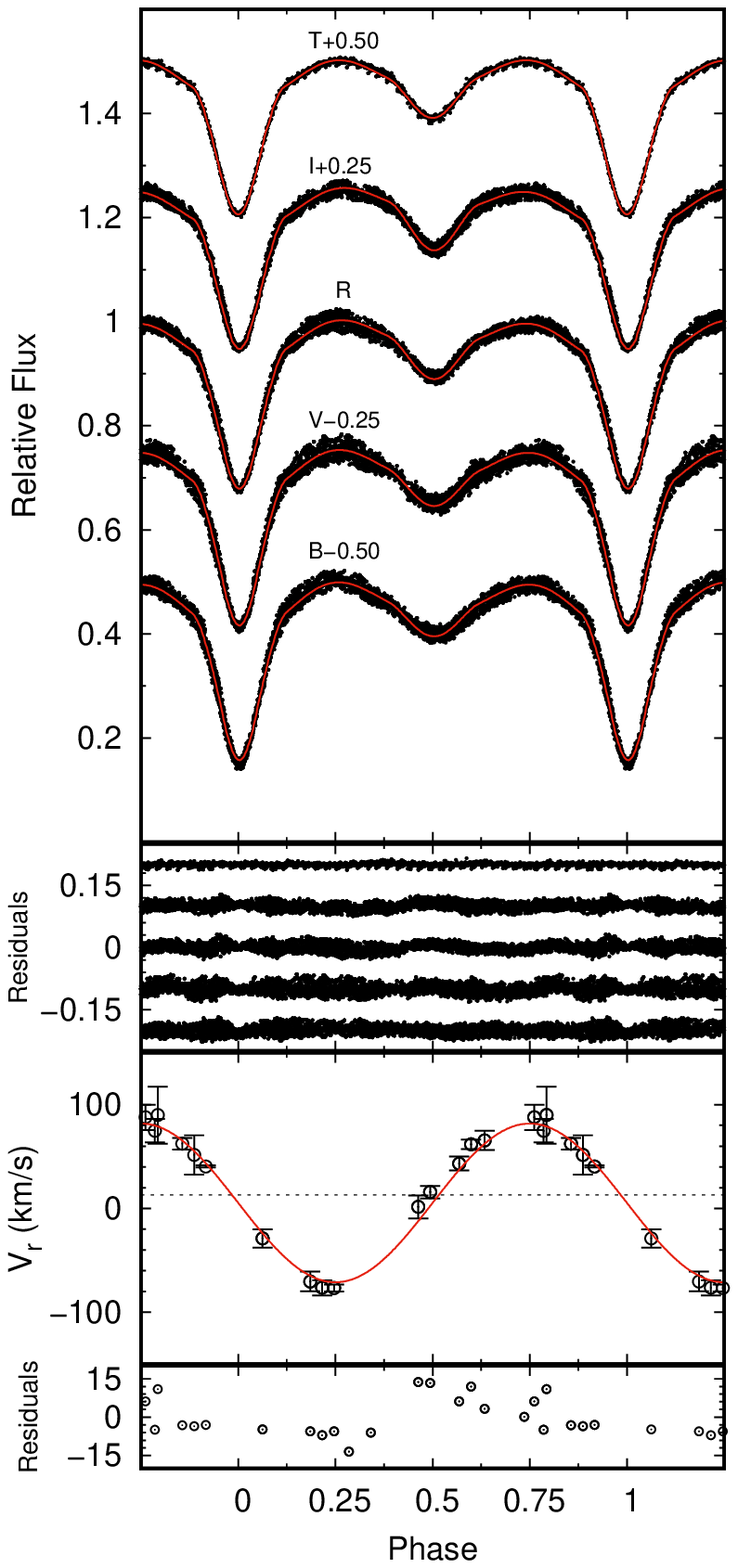}
\FigCap{Same as Fig.~1, but for {GQ~Dra}.}
\label{figlcs}
\end{figure}

\MakeTable{lcccc}{12.5cm}{Absolute parameters of the systems. The standard errors corresponding to the last digits are given in parentheses.}
{\hline
Parameter                &\multicolumn{2}{c}{V1241~Tau} &\multicolumn{2}{c}{GQ~Dra}    \\
\hline
& \multicolumn{1}{c}{Pri.}&\multicolumn{1}{c}{Sec.}&\multicolumn{1}{c}{Pri.}&\multicolumn{1}{c}{Sec.}\\
\hline
M (M$_{\odot}$)          & 1.8(4)  & 0.8(3) &   1.8(7)&0.6(2) \\
R (R$_{\odot}$)          & 1.8(1)  & 1.7(1) &   1.9(2) &1.3(1) \\
L (L$_{\odot}$)		     & 7(1) & 1.4(2) &  20(4) &1.1(2)\\
%M$_{bol}$ ($^m$)         &2.6(5.1) & 4.6(4.8) &1.3(1.2) &4.5(1.5) & 1.2(3) & 4.6(1.1) \\
$a$ (R$_{\odot}$)        & \multicolumn{2}{c}{5.1(4)} & \multicolumn{2}{c}{4.7(5)}  \\
\hline}

\section{Orbital Period Study}

The $O-C$ diagram analyses were made using the least squares method with statistical weights (w=1 for visual observations, 1-3 for photographic, and 10 for CCD and photoelectric observations) using a MATLAB code. Details about this code and the set of parameters used can be found in Zasche~{\it et al.} (2009). The $O-C$ diagrams of the systems were made using the times of minima calculated in this study (Table~2), and past minima timings and ephemerides from the ``$O-C$ gateway'' database\footnote{http://var2.astro.cz/ocgate/}. The parameters for each orbital period modulating mechanism were calculated using the InPeVEB software (Liakos 2015). The parameters of the Light-Time Effect (LITE), Applegate, and mass transfer mechanisms are strongly depended on the absolute parameters of the systems\textsc{'} components as well as on their photometric model.

\subsection{V1241~Tau}

The $O-C$ data points of V1241~Tau show cyclic variation, that potentially can be attributed either to the presence of a third body (LITE) or to the quadrupole moment variation of the secondary component (Applegate mechanism), according to the criterion of Lanza and Rodon{\`o} (2002).

Following the formalism of  Liakos and Niarchos (2013), i.e. assuming that the tertiary component is a main-sequence star in coplanar orbit with the binary system, and for a minimum mass of M$_{3,~min}$=0.3~M$_{\odot}$, its luminosity contribution is 0.18\%. This low luminosity ratio justifies the absence of the third light in our photometric model (see Sec.~3). The stability of the third body\textsc{'}s orbit was tested using the Harrington\textsc{'}s criterion (Harrington 1977) and was found stable. The results of the analysis are shown in Table~6 and the LITE curve is plotted in Fig.~3.

\begin{figure}[htb]
\centering
\includegraphics[scale=0.55]{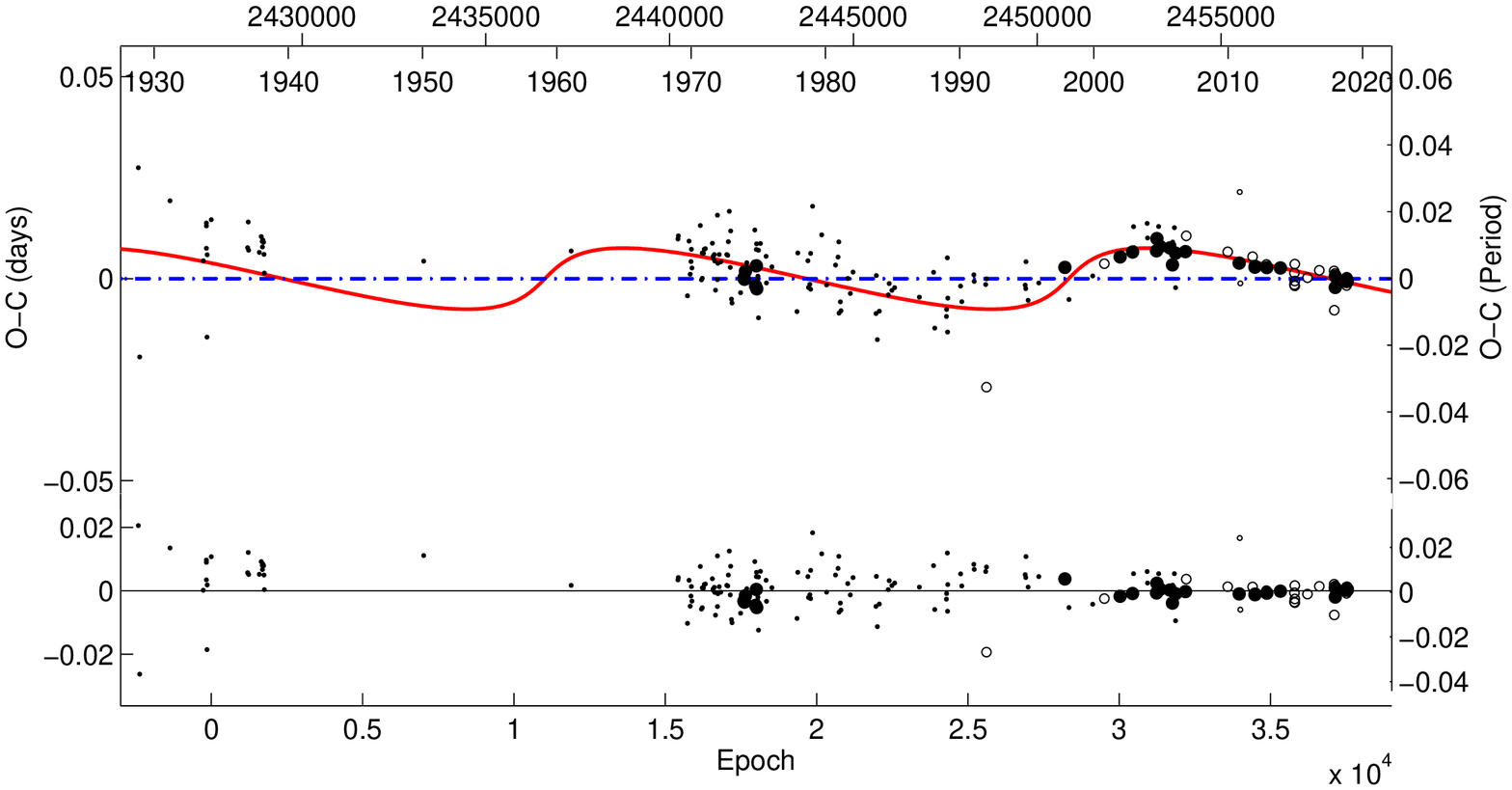}
\FigCap{$O-C$ diagram of V1241~Tau. The solid line (red in colored version) represents the LITE curve. The residuals are also shown at the bottom of the figure. The bigger the symbol the bigger the weight assigned. Filled circles denote the primary, while open circles refer to the secondary minima. Crosses represent either photographic or visual data.}
\label{figlcs}
\end{figure}

\MakeTable{lcc}{12.5cm}{O-C analysis results of the systems for LITE and Applegate mechanism.}
{\hline
Parameter               & V1241~Tau & GQ~Dra      \\
\hline
\\[-1em]
\multicolumn{3}{c}{Eclipsing binary}\\
\hline
T$_0$ (HJD) &	2427531.672(1)	& 2448500.940(3) \\
P (d)       &	0.8232711(1)	& 0.7659029(2)	\\
\hline
\\[-1em]
\multicolumn{3}{c}{Third body - LITE}\\
\hline
P$_3$ (yr)      &	39(1)		& 27(2)	\\
T$_{0,3}$ (HJD) &	2450853(503)	& 2448840(724)	\\
A$_3$ (d)       &	0.0076(8)	& 0.0076(7)	\\
$\Omega_3$ (deg) &	0(7)		& 149(21)		\\
e$_{3}$ 	    &	0.6(2)		& 0.5(3)	\\
$f$(m$_{3}$) (M$_\odot$)& 	0.0032(1)	& 0.0044(1)	\\
M$_{3,min}$ (M$_\odot$ )&	0.30(1)		& 0.32(8)	\\
$\Sigma$res$^2$	        &	0.013		& 0.0007	\\
\hline
\\[-1em]
\multicolumn{3}{c}{Quadrupole moment variation for the secondary comp.}\\
\hline
\\[-1em]
$\Delta Q$~(gr~cm$^2$) ($\times$10$^{50}$)	&	0.8(3)		& 0.7(2)	\\
\hline}

\subsection{{GQ~Dra}}

The $O-C$ analysis of the system GQ Dra, on the other hand, proceeded by examining three possible orbital period modulating mechanisms: (i) mass transfer from the secondary component to the primary, (ii) presence of a third body, and (iii) magnetic braking. %The former mechanism appears as an upward parabola fitting function in the O-C diagram while the latter ones as cyclic variations. 

We tested all the above hypotheses and we found that a cyclic period change fits better the data. The Applegate mechanism was tested and found that the orbital period changes cannot be explained by the quadrupole moment variation of the secondary component according to the Lanza and Rodon\`{o} (2002) criterion. Therefore, the existence of a third body orbiting the binary system (LITE mechanism) can be plausibly adopted as the most possible modulating mechanism of the system's orbital period. According to our results for this mechanism and following the same formalism as in the case of V1241~Tau, the minimum mass of the tertiary components is 0.32~M$_{\odot}$. The light contribution of this body is calculated as 0.09\%, which justifies its non-appearance in photometric model. The Harrington\textsc{'}s criterion (Harrington 1977) for this system results in a stable orbit. The results of the analysis are presented in Table~6 and the LITE curve is plotted in Fig.~4.

%Finally, although the LITE curve fits better the $O-C$ points of GQ~Dra in terms of $\Sigma (res)^2$, we note that this may be apparent, because it contains more free parameters in comparison with the parabola. We conclude that new minima timings are needed within the following decade(s) in order to verify the orbital period modulating mechanism of this system.

\begin{figure}[htb]
\centering
\includegraphics[scale=0.6]{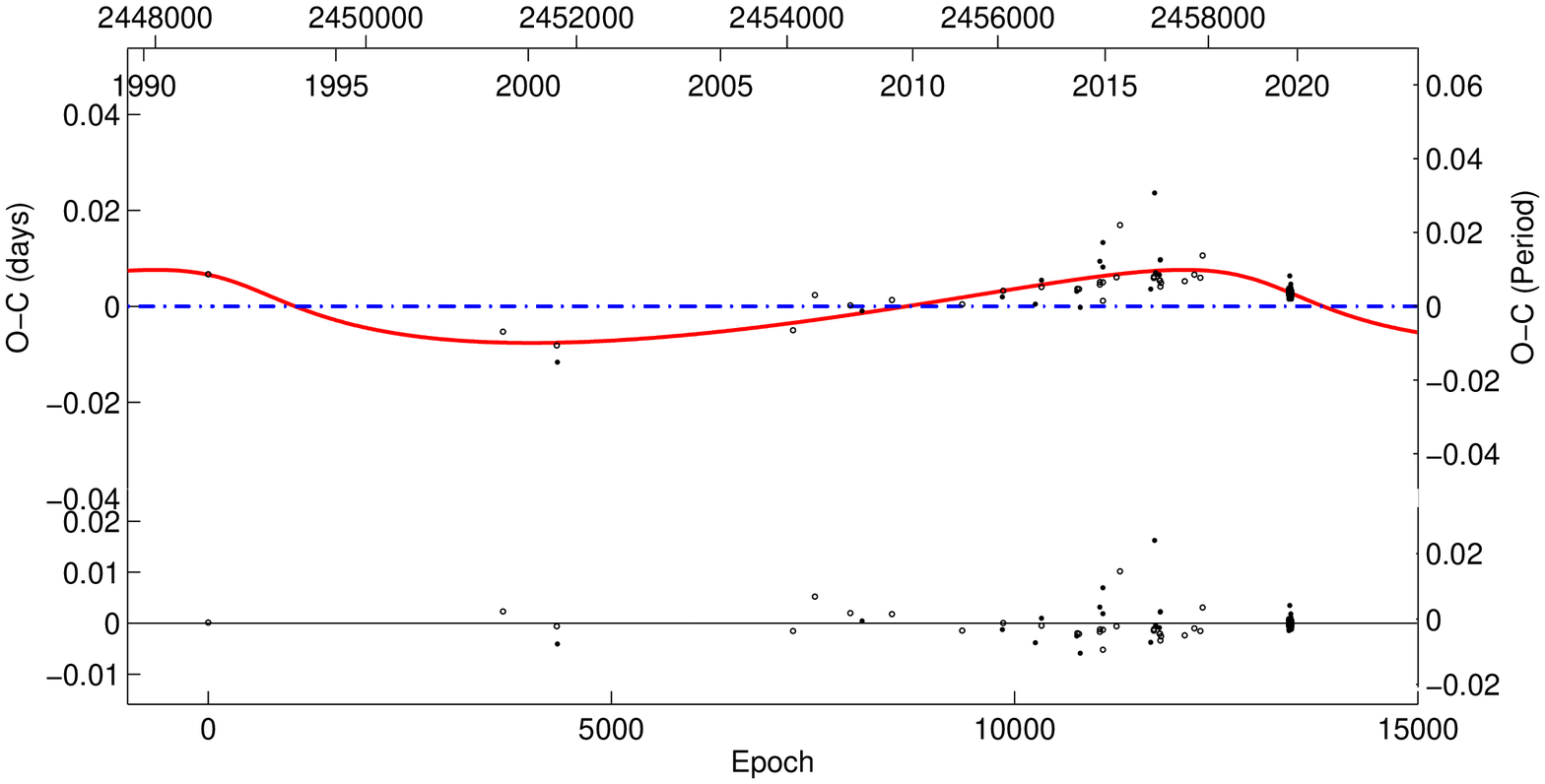}
\FigCap{$O-C$ diagram of GQ~Dra. The solid curve represents the LITE mechanism assumption. The residuals are also given at very bottom of the figure. Symbols have the same meaning as in Fig.~3.}
\label{figlcs}
\end{figure}

\section{Search for Pulsations}

Higher signal-to-noise ratios of pulsation frequencies are expected in $B$ filter for  $\delta$~Sct stars (for instance, Arentoft~{\it et al.} (2007) remarked that the ratio of $\Delta B / \Delta I $ is about 0.5). Therefore, we first analysed the $B$ filter residual data points of the out-of-eclipse phases (i.e. 0.1-0.4 and 0.6-0.9), yielded after the binary model removal, in order to search for possible pulsations. The {\tt PERIOD04} software (Lenz and Breger 2005), that is based on the classical Fourier analysis, was used for extracting the main frequencies and for calculating the amplitude spectra in the range 0-80~d$^{-1}$ for both systems. Briefly, we applied the following method: After the first frequency computation, the software searches for another one. The search is repeated after prewhitening all the previously detected frequencies. The search continues until the detected frequency has a signal-to-noise ratio (S/N)=4, which is the critical limit of the software. We calculated average noise values between 55-75~d$^{-1}$ for V1241~Tau and 60-80~d$^{-1}$ for GQ~Dra, in which all peaks in the amplitude spectrum are below the significance level. The average noises are 0.44~mmag and 0.37~mmag for V1241~Tau and GQ~Dra, respectively. The signal-to-noise ratio for each detected frequency was then adopted as the ratio of the derived amplitude to the average noise. 

We also applied frequency analyses to the out-of-eclipse residuals yielded by subtracting the binary model from TESS light curves of both systems. Very low significance levels were achieved (Figs 7 and 9) during the analyses thanks to the high quality photometric observations of the satellite.

\subsection{V1241~Tau}

The $B$ residual data points of the system were analyzed  and resulted in ten possible frequencies, which are given in Table~7. We remark that the frequencies $f_{1}$, $f_{2}$ and $f_{3}$, which are smaller than 5~d$^{-1}$, probably rise from the removal of the binary model, atmospheric conditions and observational drifts. All the other frequencies are between the valid interval for  $\delta$~Sct regime (i.e. 5-80~d$^{-1}$, Breger 2000). However, it is worth to point out that six of those frequencies may also be combination frequencies such as $f_{4} = f_{1} + f_{2} + 2f_{3}$, $f_{5} = f_{2} + f_{4} + f_{1}$, $f_{7} = f_{1} + f_{4} - f_{2}$, $f_{8} = f_{4} + 2f_{2} + 2f_{6}$, $f_{9} = f_{8} - f_{1}$, $f_{10} = f_{8} - f_{2}$ by taking into account the frequency resolution (0.121). $f_{6}$ is also in $\delta$~Sct type frequency interval, however, we concluded that this frequency is spurious since no satisfactory fit on observations was achieved. It may have occurred during the removal process of the binary model. The residuals ($\chi^2$) for the last calculated fit was found to be 0.007. Additionally, we could not find the two frequencies (6.077 and 7.288~d$^{-1}$) suggested by Sarma and Abhyankar (1979). Keeping in mind the negative results of several previous studies, it can be concluded that the resulting frequencies of Sarma and Abhyankar (1979) (fifth and sixth harmonics of the orbital period) are probably spurious and they arise from their rectification process. The spectral window and the amplitude spectrum of our residual data are shown in Fig.~5. In conclusion, the $\delta$~Sct--type behaviour of the primary component is rather ambiguous and cannot be clearly verified and new observations with higher photometric accuracy are needed to justify the result.

The residuals from TESS light curve was analysed in the method similar to previous analysis of $B$ light curve and no any frequency that can be attribute to pulsation can be found. The results are listed in Table~8. All derived frequencies are lower than 5~d$^{-1}$ and they probably are harmonics and combinations such as: $f_{1} = 4f_{orb}$, $f_{2} = 2f_{1}$, $f_{3} = f_{2}$, $f_{4} = 2f_{orb}$, $f_{5} = 2f_{4}$, $f_{6} = 2f_{2}$, $f_{7} = f_{6}$,  $f_{8} = f_{1} + f_{7}$, $f_{9} = f_{4} + f_{5}$.  The resulting residuals are 0.001575. Additionally, the frequency 33.721~d$^{-1}$, found in the analysis of $B$ residuals, could not be confirmed by the analysis of the TESS data.
The spectral window derived from the analysis is plotted in Fig.~6 with the amplitude spectrum.

\MakeTable{lcccc}{12.5cm}{Frequency analysis results of $B$ residuals for both systems. The quantities $f$, $A$, $\phi$ and S/N refer to the frequency, amplitude, phase, and signal to noise ratio, respectively. The least square uncertainties concern the last digit(s) and they are given in parentheses.}
{\hline
& $f$ (d$^{-1}$) & $A$ (mag) & $\phi$ (2$\pi$ rad) &  S/N \\
\hline
\multicolumn{5}{c}{V1241~Tau}\\
\hline
$f_{1}$&2.111(3)&0.0104(5)&0.922(7)&24\\
$f_{2}$&0.497(4)&0.0075(5)&0.607(10)&17\\
$f_{3}$&4.009(5)&0.0058(5)&0.112(13)&13\\
$f_{4}$&10.608(8)&0.0037(5)&0.588(20)&8\\
$f_{5}$&8.97(12)&0.0026(5)&0.889(28)&6\\
$f_{6}$&33.721(14)&0.0022(5)&0.265(34)&5\\
$f_{7}$&12.330(14)&0.0023(5)&0.497(32)&5\\
$f_{8}$&78.930(11)&0.0029(5)&0.506(25)&7\\
$f_{9}$&76.668(14)&0.0022(5)&0.439(33)&5\\
$f_{10}$&78.342(15)&0.0020(5)&0.236(36)&5\\
\hline
\multicolumn{5}{c}{GQ~Dra}\\
\hline
$f_{1}$&1.237(2)&0.0044(3)&0.423(10)&12\\
$f_{2}$&0.498(2)&0.0034(3)&0.707(13)&9\\
$f_{3}$&3.920(2)&0.0049(3)&0.793(9)&13\\
$f_{4}$&3.014(1)&0.0055(3)&0.576(8)&15\\
$f_{5}$&0.550(2)&0.0043(3)&0.177(10)&12\\
$f_{6}$&18.576(4)&0.0023(3)&0.592(20)&6\\
$f_{7}$&29.374(4)&0.0022(3)&0.559(21)&6\\
$f_{8}$&0.115(3)&0.0028(3)&0.617(16)&8\\
$f_{9}$&10.196(4)&0.0021(3)&0.804(21)&6\\
$f_{10}$&38.457(5)&0.0017(3)&0.736(26)&5\\
$f_{11}$&12.879(5)&0.0015(3)&0.926(30)&4\\
\hline}

\MakeTable{lcccc}{12.5cm}{Same as Table~7, but for the residuals from the TESS light curves.}
{\hline
& $f$ (d$^{-1}$) & $A$ (mag) & $\phi$ (2$\pi$ rad) &  S/N \\
\hline
\multicolumn{5}{c}{V1241~Tau}\\
\hline
$f_{1}$&0.03105(7)&0.00742(2)&0.1194(5)&43 \\
$f_{2}$&0.11643(21)&0.00242(2)&0.4822(16)&13 \\
$f_{3}$&0.14553(28)&0.00186(2)&0.3609(21)&11 \\
$f_{4}$&1.19727(36)&0.00145(2)&0.7335(26)&9 \\
$f_{5}$&2.42946(29)&0.00179(2)&0.8135(21)&15 \\
$f_{6}$&0.27555(53)&0.00097(2)&0.5554(39)&5 \\
$f_{7}$&0.30465(67)&0.00077(2)&0.4070(50)&5 \\
$f_{8}$&0.39003(73)&0.00071(2)&0.8177(54)&5 \\
$f_{9}$&3.64419(78)&0.00066(2)&0.5559(58)&6 \\
\hline
\multicolumn{5}{c}{GQ~Dra}\\
\hline
$f_{1}$&0.2629(9)&0.0027(1)&0.669(7)&18 \\
$f_{2}$&0.0261(9)&0.0030(1)&0.797(6)&19 \\
$f_{3}$&5.2235(13)&0.0019(1)&0.260(10)&14 \\
$f_{4}$&1.3064(11)&0.0022(1)&0.741(8)&13 \\
$f_{5}$&0.1565(23)&0.0011(1)&0.404(16)&7 \\
$f_{6}$&0.4916(28)&0.0009(1)&0.970(20)&6 \\
$f_{7}$&0.4054(30)&0.0008(1)&0.050(22)&6 \\
$f_{8}$&6.5279(39)&0.0007(1)&0.317(28)&5 \\
$f_{9}$&0.0662(37)&0.0007(1)&0.941(26)&4 \\
\hline}

\begin{figure}[htb]
\centering
\includegraphics{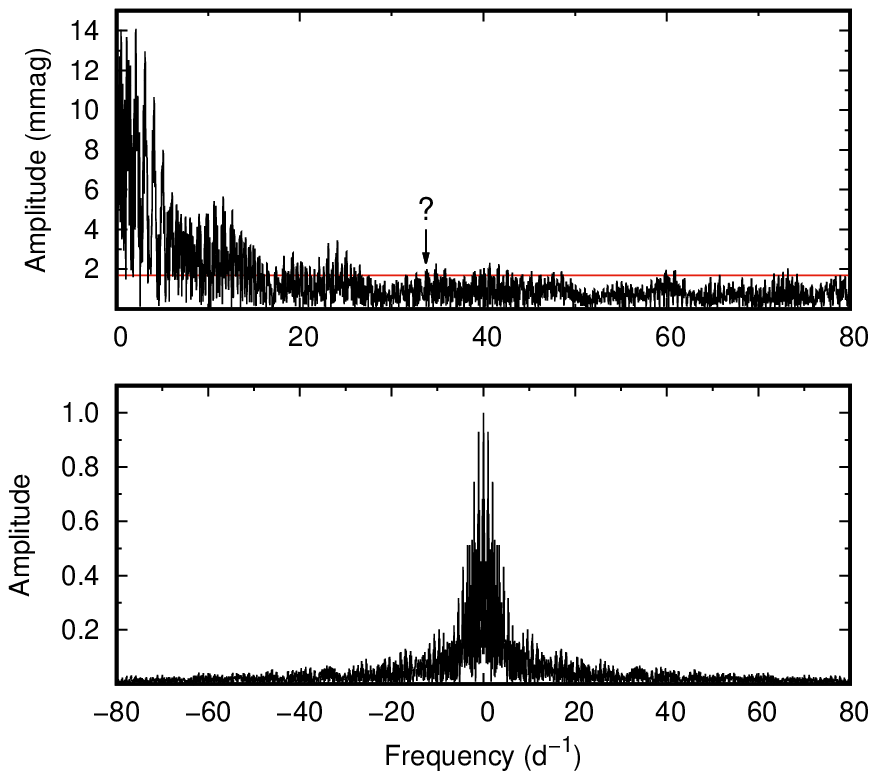}
\FigCap{Amplitude spectra (top) and spectral window (bottom) of the residual data of {V1241~Tau}. The horizontal line indicates the significance level. The frequency f$_6$, which might be a potential genuine frequency, is also indicated. For details see text.}
\label{figlcs}
\end{figure}

\begin{figure}[htb]
\centering
\includegraphics{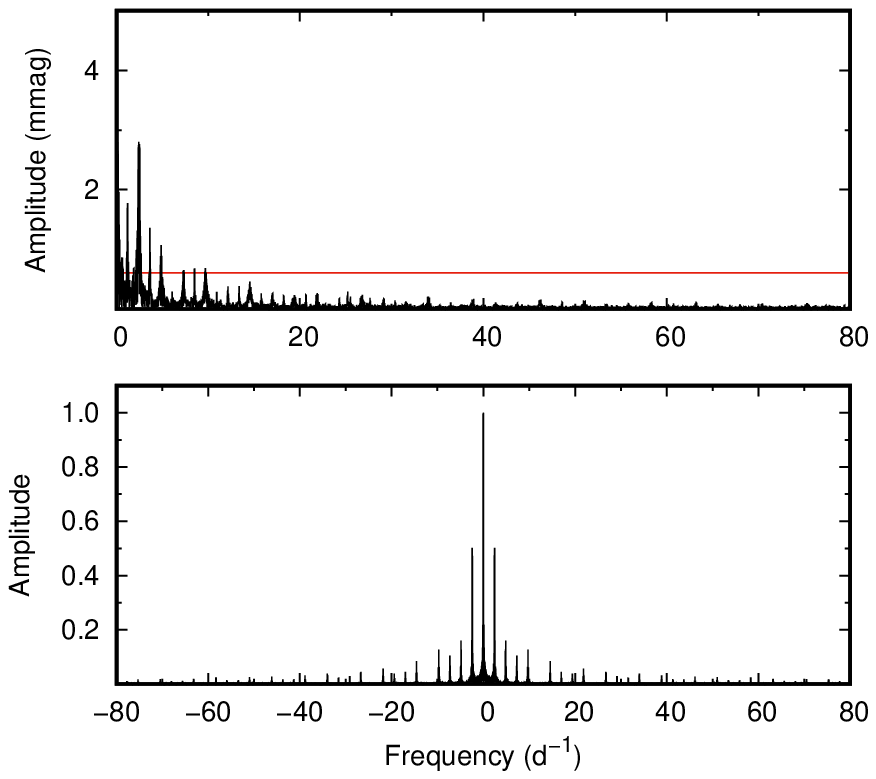}
\FigCap{Same as Fig.~5, but for the residuals from the TESS light curve.}
\label{figlcs}
\end{figure}

\subsection{{GQ~Dra}}

For the analysis of this system we used 1817 data points of $B$ residuals with a time resolution of $\sim$25~s. We detected 11 frequencies with S/N$>$4, and they are listed in Table~7. The frequencies below 5~d$^{-1}$ are probably the consequence of small fluctuations of the nightly mean values. $f_{1}$=1.237~d$^{-1}$ is very close to the frequency of the orbital period ($f_{orb} = 1.306~d^{-1}$). $f_{6}$=18.576~d$^{-1}$ is a genuine frequency and lies between 5-80~d$^{-1}$, which is typical interval for $\delta$~Sct type pulsators (Breger 2000). In addition, the temperature of the primary component (see Sec.~2 and 3) is well inside the temperature range of the $\delta$~Sct stars (A-F spectral types). Therefore, it can be concluded that this star is the oscillating component of this system. Most of the other detected frequencies also indicate oscillations within the $\delta$~Sct regime. However, the possibility of being combinations must be noted; $f_{3} = 3f_{orb} = 2f_{1} + 3f_{2}$, $f_{4} = f_{2} + 2f_{1}$, $f_{5} = f_{2}$, $f_{7} = f_{4} + f_{6} + 2f_{3}$, $f_{8} = f_{5} - f_{2}$, $f_{9} = f_{7} - f_{5} - f_{6}$, $f_{10} = f_{1} + 2f_{6}$ and $f_{11} = f_{3} + f_{9} - f_{1}$.  The combination frequencies were calculated using the frequency resolution value 0.078, which is very close to the Rayleigh criterion. The residuals of the solution were derived as 0.008. The amplitude spectrum, the spectral window plot, and the Fourier fit on the data of a selected night are shown in Fig.~7.

We also analysed the TESS light curve residuals in order to confirm the genuine frequency found in $B$ filter data. However, the TESS data consist of 30 min cadence and corresponds to very low Nyquist frequency ($\sim$24~d$^{-1}$) which is not convenient to compare it with $B$-filter frequency analysis. Nevertheless, nine frequencies were found to have signal-to-noise higher than 4.0. Seven of them are possible harmonics and combinations, while the other two frequencies ($f_{1}$ and $f_{2}$) seem to appear as a consequence of the removal of the binary model. Harmonics and combination frequencies are $f_{3} = 4f_{orb}$, $f_{4} = f_{orb}$, $f_{5} = 4f_{2}$, $f_{6} = 2f_{1}$, $f_{7} = f_{1} + f_{5}$, $f_{8} = f_{3} + f_{4}$, $f_{9} = f_{2}$. The residuals of the analysis were estimated as 0.002. The results are given in Table~8 and Fig.~8.

\begin{figure}[htb]
\centering
\includegraphics{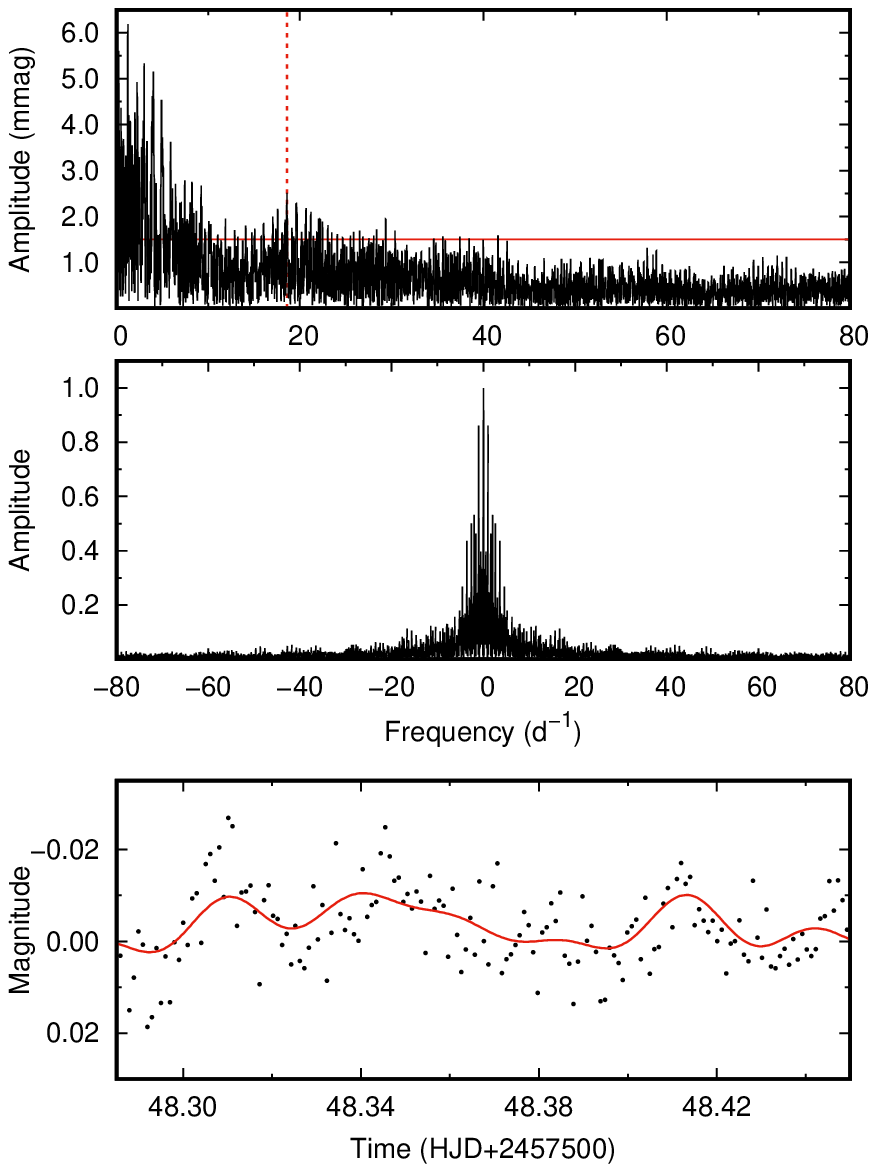}
\FigCap{Amplitude spectrum (top), spectral window (middle) and Fourier fit on selected $B$--filter residual data (bottom) of {GQ~Dra}. In the amplitude spectrum, the dashed lines (red in colored version) denote the genuine frequency, 18.576~d$^{-1}$, while the horizontal line indicates the significance level.}
\label{figlcs}
\end{figure}

\begin{figure}[htb]
\centering
\includegraphics{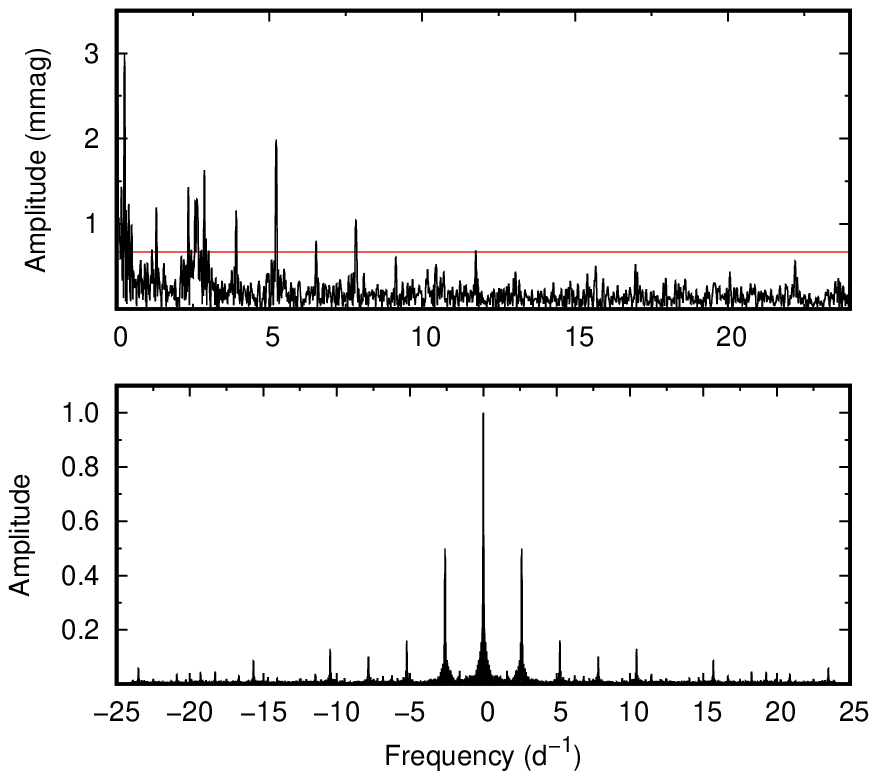}
\FigCap{Same as top and middle panel of Fig.~7, but for the residuals from the TESS light curves.}
\label{figlcs}
\end{figure}

\section{Evolution}

The locations of the components of the systems in the Hertzsprung--Russell diagram and the mass--radius plane were compared to the components of some well--known Algol--type binaries Ibano{\v{g}}lu~{\it et al.} (2006). As seen in Fig.~9, the components of both systems are in accordance with the other Algols.

\begin{figure}[htb]
\centering
\includegraphics{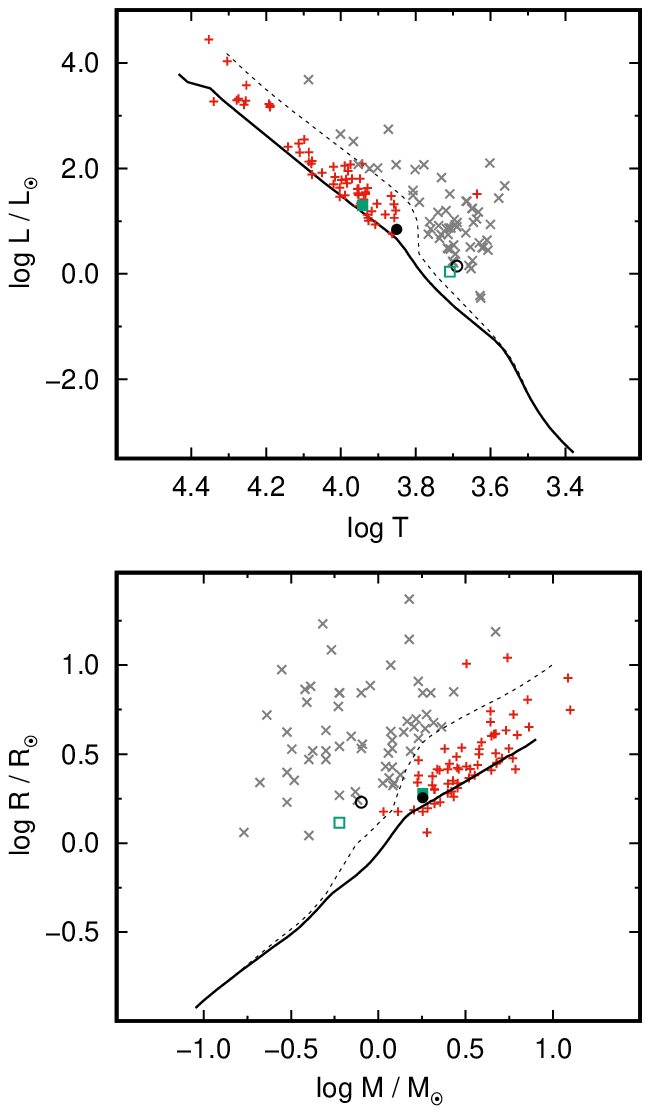}
\FigCap{Top panel: Location of the components of the two studied systems in the Hertzsprung--Russell diagram. Solid and dashed lines represent the zero and the terminal age main--sequence. Bottom panel: Location of the components in the mass--radius plane along with other components of Algol type binary systems. Plus signs denote the primary components of Algols, while the secondaries are marked with crosses. Filled and open circles refer to the primary and secondary components of {V1241~Tau}, while filled and open squares (green in colored version) denote the respective components of {GQ~Dra}. The parameters of the Algols were taken from Ibano{\v{g}}lu~{\it et al.} (2006), while ZAMS and TAMS data from Bressan~{\it et al.} (2012) based on the mass fraction values of Z=0.01 and Y=0.267.}
\label{fighr}
\end{figure}

The evolutionary paths of the systems, by comparing their location with the evolutionary tracks of some binary models assuming solar abundances, were calculated and plotted in Fig.~10. The Binary  Star Evolution ({\tt BSE}) code (Hurley~{\it et al.} 2013, 2002) was used for the derivation of the evolutionary tracks with the initial values of  eccentricity e$_{i}$ between 0 and 1, orbital period of the system P$_{i}$ between 3~d and 7~d, mass of the primary component M$_{1i}$ between 1.0 M$_{\odot}$ and 10.0 M$_{\odot}$, and mass of the secondary component M$_{2i}$ between 0.1 M$_{\odot}$ and M$_{1}$ by limiting the evolution time to 15~Gyr. The evolutionary tracks, plotted in Fig.~10, were selected as the closest among thousands generated by the code. That is e$_{i}$=0.6, $P_{i}$=2.6, M$_{1i}$=1.1~M$_{\odot}$, M$_{2i}$=0.8~M$_{\odot}$ for {V1241~Tau} and e$_{i}$=0.7, $P_{i}$=3.4, M$_{1i}$=1.7~M$_{\odot}$, M$_{2i}$=0.8~M$_{\odot}$ for {GQ~Dra}. It should be noted that the present primaries are close to the evolutionary tracks of the originally secondary components. This is also valid for the present secondaries.

Our results agree with the theory of binary star evolution. That is the semi-detached systems evolves from detached binaries. Since the evolution of the massive component is faster it fills its Roche lobe and transfer mass to originally less massive, present more massive, companion (Ritter 1996). This process results with the occurrence of mass ratio reversal and forming a semi-detached system.

\begin{figure}[htb]
\centering
\includegraphics{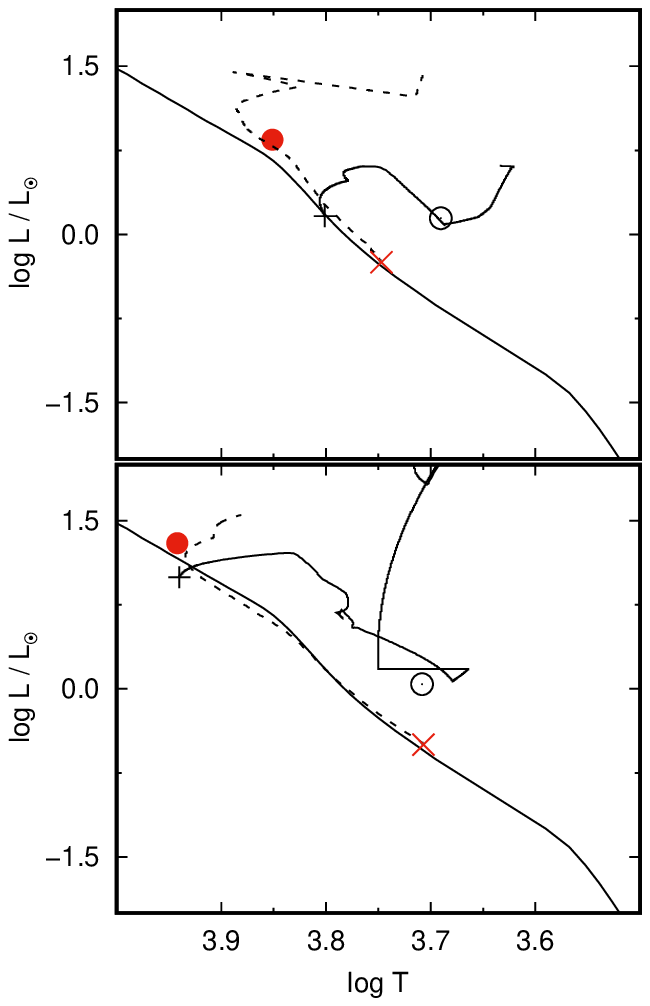}
\FigCap{Locations of the components of {V1241~Tau} (top panel) and {GQ~Dra} (bottom panel) in the Hertzsprung--Russell diagram along with the evolutionary tracks of the BSE models (for details see text). The filled circles (red in colored version) represent the present primaries, while the open circles refer to the present secondaries. The solid and dashed lines show the evolutionary tracks of the initial primary and secondary components, respectively. Plus and cross (red in colored version) signs denote the zero age positions of the initial primary and secondary components.}
\label{figlcs}
\end{figure}

\section{Summary and Conclusions}

We presented the analyses of multi--colour light and radial velocities curves, orbital period study,  and frequency search for the binary systems {V1241~Tau} and {GQ~Dra}. The TESS light curves of the systems were also incorporated into the analyses. The results showed that both systems are semi--detached binaries; a hot primary and a cooler secondary component which fills its Roche lobe. We combined our photometric and spectroscopic data to estimate the physical parameters of the systems, which were also compared to other similar ones. Moreover, the evolutionary tracks of the systems were also calculated.

Our results for {V1241~Tau} show slight difference from those of previous studies. The orbital inclination we found confirms that of Yang~{\it et al.} (2012) and it is lower than the value of  Giuricin and Mardirossian (1981) (84$\si{\degree}$.9) and Russo and Milano (1983) (86$\si{\degree}$.4). The derived mass ratio value of the system is also quite different than that of Arentoft~{\it et al.} (2004), who suggested two different values (0.36 and 0.75) with different solutions, while it is supported by the results of Giuricin and Mardirossian (1981) and Yang~{\it et al.} (2012), who derived a mass ratio as 0.5 and 0.545, respectively. However, it should be noted that our solution is the first that is based on high quality photometric data as well as on radial velocities measurements. Therefore, we claim that the present results can be considered as more accurate.

The system {V1241~Tau} was indicated as an oEA star by a few previous studies. Our Fourier analysis shows that its residual light curve has no exact trace of pulsational properties beyond orbital frequency, a spurious frequency, low--frequency artifact effects and combination frequencies. The frequencies which were attributed as possible oscillations in the literature Sarma and Abhyankar (1979) are somehow spurious and may related to the orbital period. According to the $O-C$ diagram analysis results, a tertiary component with a minimum mass of 0.3~M$_{\odot}$ orbits the binary system and modulates its orbital period.

The resulting values for {GQ~Dra} also differ from the  only analysis in literature, i.e. UV light curve analysis by Qian~{\it et al.} (2015).  Our resulting mass ratio and surface potential values are larger, while we found lower inclination and temperature value of the secondary component comparing to theirs. Similarly to V1241 Tau, our results should be considered as more accurate since they are based on modern high quality photometric observations accompanied by high resolution spectroscopy.

We can plausibly conclude that the primary of {GQ~Dra} is a pulsating star of $\delta$~Sct type with at least one genuine frequency of 18.58~d$^{-1}$ based on our $B$ residuals analysis although we could not confirm this result by the frequency analysis applied to the TESS light curve between 0 and 24~d$^{-1}$ interval. Moreover, given that the system has a conventional semidetached geometrical configuration (see Sec.~3) and according to the definition of Mkrtichian~{\it et al.} (2002), from now on GQ Dra can be considered as an oEA type system. %Due to absence of enough past minima timings, the orbital period modulating mechanism cannot be verified yet. Both LITE and mass transfer from the less to the more massive component explain well the $O-C$ variations.
According to the $O-C$ analysis, the most possible orbital period modulating mechanism is the LITE. However, due to the absence of past minima timings, future ones are needed in order to confirm the present results.

\begin{sloppypar}
\Acknow{CU acknowledges the research grant No. R00125 administered by the office of the DVC Research, Development and Innovation at Botswana International University of Science and Technology (BIUST). Southern Cross Astronomy Research Group would like to thank BIUST for the support from grant No. R00125. This research has made use of the {SIMBAD} database, operated as CDS, Strasbourg, France and the ``$O-C$ gateway'' database (http://var2.astro.cz/ocgate/). This paper includes data collected by the TESS mission, which are publicly available from the Mikulski Archive for Space Telescopes (MAST). The light curve data used in this study are available upon request through direct communication with the authors.}
\end{sloppypar}

\end{document}